\documentclass[10pt,journal,compsoc]{IEEEtran}
\usepackage{multirow}
\usepackage{etoolbox}
\makeatletter
\@ifundefined{color@begingroup}%
{\let\color@begingroup\relax
\let\color@endgroup\relax}{}%
\def\fix@ieeecolor@hbox#1{%
\hbox{\color@begingroup#1\color@endgroup}}
\patchcmd\@makecaption{\hbox}{\fix@ieeecolor@hbox}{}{\FAILED}
\patchcmd\@makecaption{\hbox}{\fix@ieeecolor@hbox}{}{\FAILED}
\usepackage{amsmath,amssymb,amsfonts}
\usepackage{algorithmic}
\usepackage{graphicx}
\usepackage{textcomp}
\usepackage{arydshln}
\usepackage{booktabs}
\usepackage{hyperref}

\def\BibTeX{{\rm B\kern-.05em{\sc i\kern-.025em b}\kern-.08em
    T\kern-.1667em\lower.7ex\hbox{E}\kern-.125emX}}
    
\usepackage[style=ieee,maxnames=6, minnames=1]{biblatex}
\begin{document}
%


\title{FSC-loss: A Frequency-domain Structure Consistency Learning Approach for Signal Data Recovery and Reconstruction}

\author{Liwen Zhang, Zhaoji Miao, Fan Yang, Gen Shi, Jie He, Yu An, Hui Hui and Jie Tian, \IEEEmembership{Fellow, IEEE}
\thanks{This work was supported in part by the National Natural Science Foundation of China under Grant 82202269, 62027901; National Key Research and Development Program for Young Scientists under Grant 2022YFC2505700. (Corresponding author: Hui Hui, Jie Tian)}
\thanks{Liwen Zhang and Zhaoji Miao contributed equally to this work.}
\thanks{Liwen Zhang, Fan Yang and Hui Hui are with the CAS Key Laboratory of Molecular Imaging, Institute of Automation, Chinese Academy of Sciences, Beijing, 100190, China (e-mail: \{zhangliwen2018, yangfan2022, hui.hui\}@ia.ac.cn).}
\thanks{Zhaoji Miao is with the School of Computer Science and Engineering, Southeast University, Nanjing, 211189, China (e-mail: 230238543@seu.edu.cn).}
\thanks{Gen Shi, Jie He, Yu An, and Jie Tian with School of Engineering Medicine and School of Biological Science and Medical Engineering, Beihang University, Beijing, 100191, China (e-mail: \{shigen, jieh,  yuan1989\}@buaa.edu.cn, tian@ieee.org).}
}


\maketitle  
\begin{abstract}


A core challenge for signal data recovery is to model the distribution of signal matrix (SM) data based on measured low-quality data in biomedical engineering of magnetic particle imaging (MPI). For acquiring the high-resolution (high-quality) SM, the number of meticulous measurements at numerous positions in the field-of-view proves time-consuming (measurement of a 37×37×37 SM takes about 32 hours). To improve reconstructed signal quality and shorten SM measurement time, existing methods explore to generating high-resolution SM based on time-saving measured low-resolution SM (a 9 × 9 × 9 SM just takes about 0.5 hours). However, previous methods show poor performance for high-frequency signal recovery in SM. To achieve a high-resolution SM recovery and shorten its acquisition time, we propose a frequency-domain structure consistency loss function and data component embedding strategy to model global and local structural information of SM. We adopt a  transformer-based network to evaluate this function and the strategy. We evaluate our methods and state-of-the-art (SOTA) methods on the two simulation datasets and four public measured SMs in Open MPI Data. The results show that our method outperforms the SOTA methods in high-frequency structural signal recovery. Additionally, our method can recover a high-resolution SM with clear high-frequency structure based on a down-sampling factor of 16 less than 15 seconds, which accelerates the acquisition time over 60 times faster than the measurement-based HR SM with the minimum error (nRMSE=0.041). Moreover, our method is applied in our three in-house MPI systems, and boost their performance for signal reconstruction.

\end{abstract}

\section{Introduction}
Signal analysis is one of core filed in modern science and engineering\cite{science2024lightwave,TKDE2023metavim}, focusing on data acquisition\cite{TKDE2022optimal}, modeling, analysis, and reconstruction of signals to extract valuable information. These techniques mainly include time-domain analysis, frequency-domain analysis, and machine learning algorithms \cite{tkde2023patient}. These techniques are widely used in different fields, such as healthcare \cite{tkdedrug2023}, signal reconstruction \cite{TMI2020reconstruction}, sociology\cite{TKDE2016,TKDE2023metavim}, and network engineering \cite{tkde2023}. In this study, we focus on biomedical engineering of Magnetic particle imaging (MPI) for frequency-domain analysis. The modality of MPI is a rapidly evolving medical imaging technique with high sensitivity for visualizing the distribution of super paramagnetic iron oxide nanoparticles (SPIONs)  \cite{8630849,mohn2022TMI,TMI_TranSMS}. Compared to traditional medical imaging modalities, such as X-rays, computed tomography (CT), and magnetic resonance imaging (MRI), MPI excels at capturing real-time dynamics and offers high sensitivity for SPION detection, which has shown great potential in clinical applications, including vascular imaging \cite{tong2023sensitive}, drug delivery \cite{huang2023deep}, and vivo cell tracking \cite{RN813}.

Currently, signal matrix (SM)-based reconstruction methods are widely-used for MPI reconstruction \cite{shi20233d,RN961yinlinSM, zhang2023mpi} due to its obvious advantage of higher resolution of reconstructed image compared to the X-space method \cite{RN849tmixspace}. In practice, accurate image reconstruction in MPI primarily relies on the repeated measurement of high-resolution SM (${\text{SM}}_{{\text{HR}}}$) \cite{mohn2022TMI}. The SM is obtained through premeasurements in a controlled environment and describes the response characteristics of magnetic nanoparticles (MNPs)  in a specific system. Although this method can provide high-quality imaging results, it requires time-consuming measurements and repeated calibration. For example, it needs to spend more than 32 hours for measuring a 37×37×37 SM \cite{3dsmr}. However, a 9 × 9 × 9 SM just takes about 0.5 hours. Therefore, to explore how to achieve accurate recovery for a ${\text{SM}}_{{\text{HR}}}$ via measuring a labor-saving low-resolution SM (${\text{SM}}_{{\text{LR}}}$) is significant for reducing calibration time for ${\text{SM}}_{{\text{HR}}}$, which has drawn much attention in recent years.
To address the issue of time-consuming ${\text{SM}}_{{\text{HR}}}$ recovery, previous studies have demonstrated that the Compressed Sensing (CS) can reduce the number of calibration scans \cite{TMI_SM,weber2015reconstructionCS}, However, CS methods rely on predefined mathematical models and fail to recover high-frequency distribution for SM recovery \cite{TMI_TranSMS}. Although traditional up-sampling methods (e.g., Bicubic) \cite{gungor2020superBic} can recover ${\text{SM}}_{{\text{HR}}}$, the methods show poor performance due to the fixed calculation based on neighboring pixel information at different locations.

In recent years, powerful deep learning methods\cite{li2024savsrAAAI,3dsmr,dong2015imageSRCNN,RN1077yinlin}, especially Transformers \cite{bi2024learning,zhang2024transformer} have brought new ideas for achieving accurate ${\text{SM}}_{{\text{HR}}}$ recovery. Baltruschat et al. proposed a novel method called 3d-SMRnet, which used a CNN to recover three-dimensional (3D) ${\text{SM}}_{{\text{HR}}}$ \cite{3dsmr}. Gungor et al. proposed a novel deep learning approach based on CNN and Transformer (named TranSMS) for ${\text{SM}}_{{\text{HR}}}$ recovery \cite{TMI_TranSMS}. The TranSMS shows good performance for accurate recovery of the ${\text{SM}}_{{\text{HR}}}$ based on sparse down-sampling SMs at scales of 2×, 4×, and 8×. However, this method requires a significant amount of time for model training (three thousand epochs). Recently, our previous study \cite{TMIshigen10189221} introduced a progressive pre-training method for 3D SM calibration in magnetic particle imaging (MPI), which proposed a transformer-based method to exploit the physical characteristics of the SM coil channel and frequency index as prior information. This approach outperforms existing methods and significantly reduces the time and labor costs of frequent SM re-calibration. Although existing methods are effective for improving accuracy of ${\text{SM}}_{{\text{HR}}}$ recovery, the performance of high-frequency signal recovery is still poor. Moreover, these methods use L1 or L2 norm as loss for pixel-level error information constraint, which do not consider structure distribution information of different frequencies of the SM, which is a core challenge for high-resolution image reconstruction.

In our study, we propose a novel learning approach of frequency structure consistency (FSC) loss function, signal component embedding strategy, and exploit a self-adaption multi-scale shifted window-based (SMSW) Transformer based on Swin-transformer backbone for signal of ${\text{SM}}_{{\text{HR}}}$ recovery, which considers signal similarity and structure as supervised information to learn the distribution of the signal in ${\text{SM}}_{{\text{HR}}}$. Our method optimizes and facilitates learning low- and high-frequency signal distribution for ${\text{SM}}_{{\text{HR}}}$ recovery so that the reconstructed images have sharp structural details. We summarize our main contributions as follows:
\begin{enumerate}
\item We propose a novel FSC-loss for ${\text{SM}}_{{\text{HR}}}$ recovery that enables model to minimize pixel differences and takes different frequency structural similarity as the guidance for learning frequency distribution of SM. We also adopt a prevalent transformer-based network ( Fig. \ref{fig1-3net}) to learn the distribution of SM.

\item We propose a new complex signal components embedding strategy for encoding SM  of their real, imaginary, and magnitude (RIM), called RIM-embedding in the following context, as three-channel input (Fig. \ref{fig1RIM}). This embedding method takes prior information of signal amplitude, phase, and frequency characteristics as guidance for learning the signal distribution.

\item Our method is evaluated on two simulation datasets and real-world Open MPI Data. The results demonstrate that our method achieves the best performance among state-of-the-art methods. Furthermore, for application test, our method is applied in three in-house MPI systems, which boosts their performance for high resolution image reconstruction and short SM calibration time. Our code will be available at \href{https://github.com/dreamenwalker/MPI\_SSIM\_mix\_loss}{https://github.com/dreamenwalker/FSC\_loss.}
\end{enumerate}

\begin{figure*}[ht!]
\centering
\includegraphics[width=\textwidth]{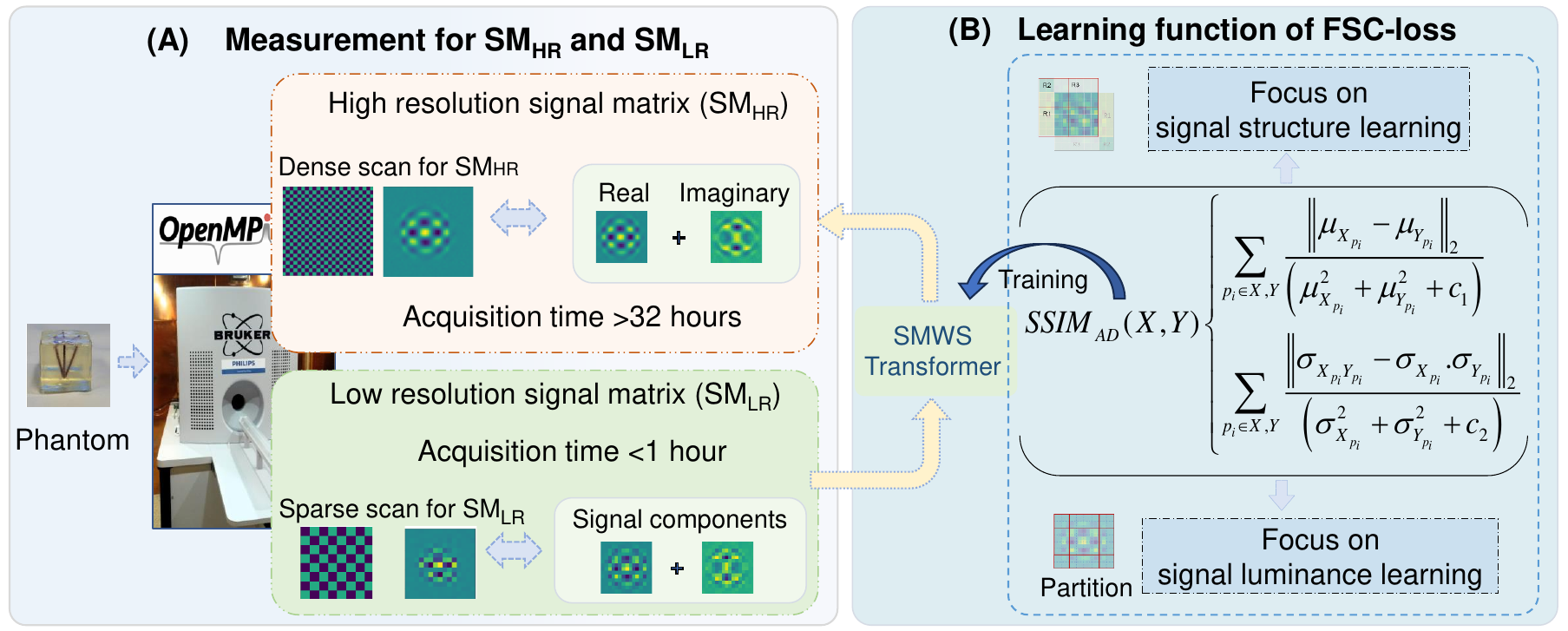}
\caption{Overview diagram of our learning approach for ${\text{SM}}_{{\text{HR}}}$ recovery. To shorten the acquisition time of the ${\text{SM}}_{{\text{HR}}}$ (more than 32h) and sharp image reconstruction, we only need to measure a ${\text{SM}}_{{\text{LR}}}$ (part A) and recover the ${\text{SM}}_{{\text{HR}}}$ by applying the method.} 
\label{fig1}
\end{figure*}
\section{Methodology}

Overview diagram of our proposed method for SM recovery is shown in Figure~\ref{fig1}. We mainly introduce four parts: 1) problem formulation for our study; 2) RIM-embedding strategy; 3) FSC loss; 4) SMSW network. 

\subsection{Problem Formulation}
MPI reconstruction based on the SM is conducted in the frequency domain. After a system receives the time-domain voltage signal $u(t)$ from measured particles, a discrete Fourier transform is applied to obtain $u(k)$. In the spatial distribution of magnetic particles within the region $\Omega$ at position $r$, with concentration $c(r)$, the relationship among the system function $s(k)$ and the signal $u(k)$ is expressed as follows:
\begin{equation}
u(k)=\int_{\Omega}s_k(r) c(r) d^3 r.
\end{equation}

When the spatial domain $\Omega$ is discretized into ${N}$ points, where $r_n \in \Omega$, ${n} \in I_N$ and $I_N$ represents an array of ${N}$ discrete points, the above equation can be transformed into:
\begin{equation}
u(k)=\sum_{{n} \in I_N} s_{k, n} c_n,
\end{equation}
where $s_{k, n} = \omega_n s_k(r)$ and $c_n = c\left(r_n\right)$, represent the value of the system function at position $r_n$ and the magnetic particle concentration at that position, respectively. $\omega_n$ represents the orthogonal weight, commonly a fixed value. The equation can be further simplified to a matrix form representing the relationship between the received voltage signal, the SM, and the particle concentration:
\begin{equation}
u=Sc,
\end{equation}
where the SM $S=\left(s_{k, n}\right)_{\mathrm{k} \in I_k, \mathrm{n} \in I_N} \in \mathbb{C}^{K \times N}$, $I_k$ is the number of row of SM. The magnetic particle concentration $c=\left(c_n\right)_{\mathrm{n} \in I_N} \in \mathbb{R}^N$, and the received voltage signal $u=\left(u_k\right)_{\mathrm{k} \in I_k} \in \mathbb{C}^K$, where $K$ represents the number of frequency points used in the reconstruction process. Each row of the SM $S$ represents the signal at different positions of the target region at the same frequency point, and each column represents the signal at different frequencies.


The number of columns represents the number of pixel points sampled in the Field of View (FOV), and the number of rows corresponds to the number of frequency points. In our work, each row of the SM is reshaped into a separate square matrix image for processing, which adopts the same approach for shape transformation of SM as in the \cite{TMI_TranSMS,3dsmr,TMIshigen10189221}. The task of SM recovery can be considered as a super-resolution task: a ${\text{SM}_{\text{HR}}}$ can be generated from a time-saving measured ${\text{SM}_{\text{LR}}}$: 
\begin{equation}
{\text{S}}{{\text{M}}_{{\text{HR}}}} = {\mathbb{F}}{\text{(S}}{{\text{M}}_{{\text{LR}}}}),
\end{equation}
 where ${\mathbb{F}}({\cdot})$ denotes a trained network. The overview is shown in Fig. \ref{fig1}.

For model training, we need to obtain ${\text{SM}_{\text{LR}}}$ and ${\text{SM}_{\text{HR}}}$. to obtain the SM [39, 40]. Main idea of this method is to place known concentration of a sample in the target imaging area for measurement. If the sample is small enough and matches the size of the discretized grid unit, the results obtained by moving the sample point by point are the discretized SM. 

Measurement-based SM Sampling: We can obtain the ${\text{SM}_{\text{HR}}}$ by employing a 3-axis linear robot to scan the field of view (FOV) at equidistant intervals on a predetermined grid. This approach allows for the acquisition of both ${\text{SM}_{\text{HR}}}$ and ${\text{SM}_{\text{LR}}}$. To minimize the travel time of the robot, the sampling trajectory of the 3D Lissajous is used. This process is similar to generating a ${\text{SM}_{\text{LR}}}$ by down-sampling every $\mathrm{n}^{\text{th}}$ pixel from a ${\text{SM}_{\text{HR}}}$. Thus, our problem formulation is how to propose an effective learning approach for generating a ${\text{SM}_{\text{HR}}}$ by a ${\text{SM}_{\text{LR}}}$. 

\begin{figure*}[h!]
\centering
\includegraphics[width=0.75\textwidth]{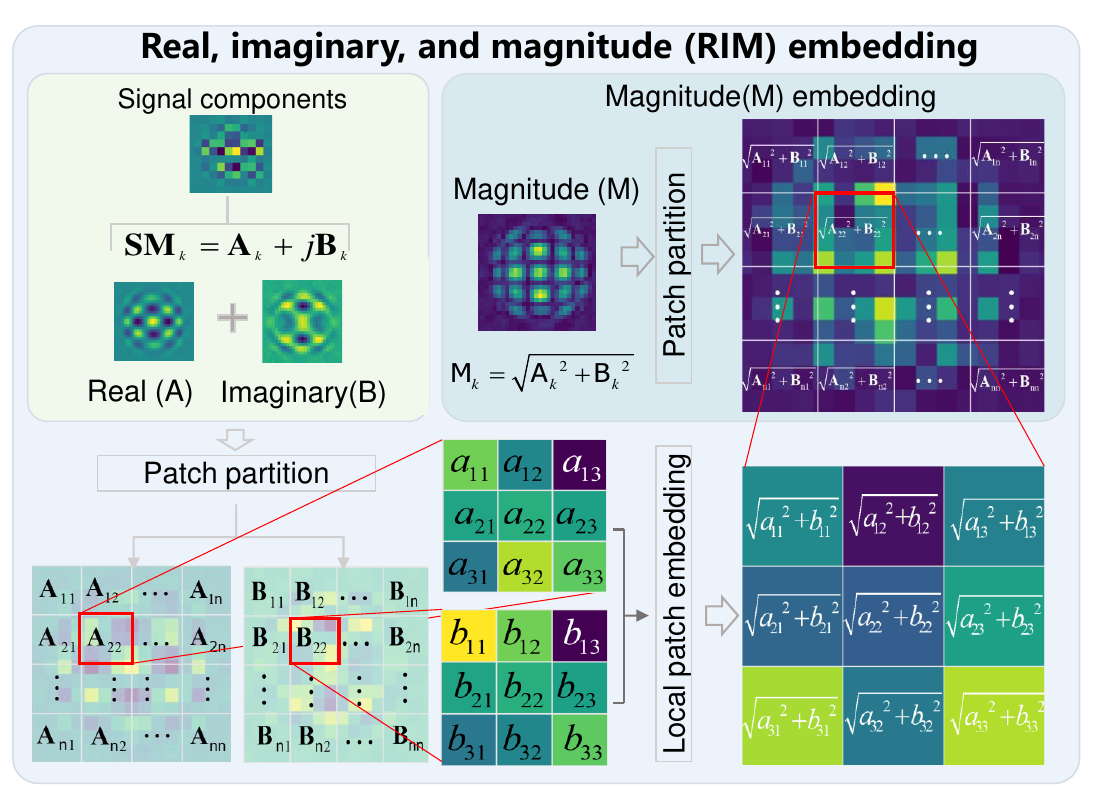}
\caption{Overview diagram of our signal component RIM-embedding strategy for ${\text{SM}}_{{\text{HR}}}$ recovery.} 
\label{fig1RIM}
\end{figure*}

\subsection{Proposed RIM-embedding}
As shown in Figure \ref{fig1RIM}, we propose a new signal component RIM-embedding strategy that encodes the SM signals into three-channel input, including the real part, the imaginary part and the magnitude. This encoding method aims to consider the signal amplitude, phase information, and signal power as input, intending to further guide model to learn signal strength, energy distribution, and frequency characteristics. More importantly, the RIM-embedding provides the additional physical prior information of the signal strength, which makes it easier to model the frequency distribution of the SM.

\subsection{Proposed FSC-Loss}
Our proposed FSC loss is inspired by SSIM loss \cite{TCI_SSIM_Zhao}. Our loss emphasizes the need for a comprehensive optimization strategy that considers both the consistency at the pixel level and the structural integrity of the SM. To easily read, we introduce representative existing loss ( $\ell_1$, $\ell_2$, SSIM)and our proposed loss.

\subsubsection{Conventional Loss ($\ell_1$ and $\ell_2$) }
As shown in work \cite{TCI_SSIM_Zhao}, to calculate the loss function $\mathcal{E}$ for SM recoery, the error for an image $P$ can be written as
\begin{equation}
\mathcal{L}^{\mathcal{E}}(P)=\frac{1}{N} \sum_{{p_i} \in P} \mathcal{E}({p_i}),
\end{equation}
where $N$ is the number of pixels $p$ in an image. For $\ell_\omega$ ($\omega$=1 or 2) is simply defined as:

\begin{equation}
{\mathcal{L}^{{\ell _\omega }}}(P) = \frac{1}{N}\sum\limits_{{p_i} \in P} | {\bf{x}}({p_i}) - {\bf{y}}({p_i}){|^\omega },
\end{equation}
where $p_i$ is the index of the pixel and $P$ is an image; ${\bf{x}}(p_i)$ and ${\bf{y}}(p_i)$ are the values of the pixels in the predicted image and the ground truth, respectively. Zhao et al. \cite{TCI_SSIM_Zhao} show that the shortcomings of these loss are summarized as: 1) regardless of the underlying local structure, only focus on pixel consistency between the predicted image and the ground truth; 2) regardless of learning luminance and contrast of images. We also experimentally demonstrate that $\ell_1$ or $\ell_2$ loss shows sub-optimal performance for frequency structure learning (in Part III).

\subsubsection{SSIM Loss}
The Structural Similarity (SSIM) loss is a widely used loss for image restoration \cite{TCI_SSIM_Zhao}. Given $x$ and $y$ are two images, the SSIM consists fo three parts: 1) Luminance function: \(l(x,y)\). The luminance comparison is based on the mean intensity of the images.  2) Constract function: \(c(x,y)\). The contrast comparison utilizes the standard deviation as a measure of image contrast. 3) Structure similarity function: \(s(x,y)\). The structural comparison is performed after normalizing the images by their respective standard deviations. This normalization removes the influence of both luminance and contrast, focusing purely on structural patterns. We define three parts as:

\begin{equation}
\begin{aligned}
& l(x, y)=\frac{2 \mu_x \mu_y+C_1}{\mu_x^2+\mu_y^2 + C_1}, \\
& c(x, y)=\frac{2 \sigma_x \sigma_y+C_2}{\sigma_x^2+\sigma_y^2+C_2}, \\
& s(x, y)=\frac{2 \sigma_{x y}+C_3}{2 \sigma_x \sigma_y+C_3}, \\
&\textit{SSIM}(x,y,p_i)=\frac{1}{N_p} \sum_{p_i \in x,y} l(p_i) \cdot c(p_i) \cdot s(p_i),
\end{aligned}
\end{equation}
where $x$ and $y$ are two images.
$N_p$ is the number of patches of $x$ and $y$; $p_i$ is the i-th patch of $x$ and $y$;
$\mu_x$ and $\mu_y$ are the means of $x$ and $y$, respectively.
$\sigma_x^2$ and $\sigma_y^2$ are the variances of $x$ and $y$, respectively.
$\sigma_{x y}$ is the covariance of $x$ and $y$.
$C_1$, $C_2$ and $C_3$ are constants to stabilize the denominator.
If we set  $C_3=C_2$, the SSIM function can be simplified as:

\begin{equation}
\textit{SSIM}(x,y,{p_i}) = \frac{1}{{{N_p}}}\sum\limits_{{p_i} \in x,y} l ({p_i}) \cdot \frac{{\left( {2{\sigma _{x{y_{{p_i}}}}} + {C_2}} \right)}}{{\left( {\sigma _{{x_{{p_i}}}}^2 + \sigma _{{y_{{p_i}}}}^2 + {C_2}} \right)}}.
\end{equation}
According to the properties of the SSIM function, each of its values is less than 1, resulting in an output range of [0,1]. Assuming the model is effective, as the similarity between images x and y increases, the $\mathrm{SSIM}(x,y)$ value gradually increases. Therefore, we typically use $1-\mathrm{SSIM}(x,y)$ as the loss function, so that the loss value decreases as the images become more similar.

In the $\mathrm{SSIM}(x,y)$ function, we can find that it is mainly based on the basic inequality: \(x^2+y^2 \geq 2xy\); from which Luminance \(l(x,y)\) and Contrast \(c(x,y)\) can be derived. The structural loss function \(s(x,y)\) can be derived through the Cauchy-Schwarz inequality \cite{altwaijry2023some}: \(\delta_x \delta_y \geq \delta_{xy}\). Based on the above mathematical principles and in the SM recovery task, we found the following shortcomings in SSIM: 1) It is based on relative difference measurement (ratio), which is sensitive to extreme values and may produce large errors at extreme values (such as very bright or very dark pixels). 2) Large variations in mean and variance throughout the image cause large fluctuations in the loss function values. 3) Since the existing SSIM function is optimized based on regions (patches) \cite{TCI_SSIM_Zhao}, it only achieves learning of overall information by focusing on local detail information step by step, making the model convergence process extremely slow.

\subsubsection{Proposed FSC-loss}

In our study, considering the above limitations of using $\ell_1$, $\ell_2$, and structural similarity (SSIM) \cite{TCI_SSIM_Zhao}. In the following, we will introduce our improvements to overcome existing shortcomings. For the Luminance $l(x,y)$ item, to keep the advantage of origin SSIM function and to address the the occurrence of large errors at extreme values, we modify the origin $l(x,y)$ and $s(x,y)$ as absolute difference Luminance ($L_{AD}(x,y)$ and absolute difference structure ($S_{AD}(x,y)$:

\begin{equation}
{L_{AD}}({p_i}) = {\Bbb E}\left( {\sum\limits_{{p_i} \in X,Y} {\frac{{{{\left\| {{\mu _{{X_{{p_i}}}}} - {\mu _{{Y_{{p_i}}}}}} \right\|}_2}}}{{\left( {\mu _{{X_{{p_i}}}}^2 + \mu _{{Y_{{p_i}}}}^2 + {c_1}} \right)}}} } \right),
\end{equation}
\begin{equation}
{S_{AD}}({p_i}) = {\Bbb E} \left( {\sum\limits_{{p_i} \in X,Y} {\frac{{{{\left\| {{\sigma _{{X_{{p_i}}}{Y_{{p_i}}}}} - {\sigma _{{X_{{p_i}}}}}.{\sigma _{{Y_{{p_i}}}}}} \right\|}_2}}}{{\left( {\sigma _{{X_{{p_i}}}}^2 + \sigma _{{Y_{{p_i}}}}^2 + {c_2}} \right)}}} } \right),
\label{equalCov}
\end{equation}
where
$i$ is the i-th patch of $X$ and $Y$;
$\mu_{{X_{p_i}}}$ is the i-th patch mean of $X$;
$\mu_{{Y_{p_i}}}$ is the i-th patch mean of $Y$;
$\sigma_{X_{p_i}}^2$ is the i-th patch variance of $X$;
$\sigma_{Y_{p_i}}^2$ is the i-th patch variance of $Y$;
${{\sigma_{X_{p_i}},\sigma_{Y_{p_i}}}}$ is the i-th patch standard deviations of $X$ and $Y$. Function $\Bbb E$($ \cdot $) can ensure that the product of various terms equals 1 and to achieve synergistic constraint, when an individual term reaches its optimal state (i.e., its value is 0), the other terms must remain effective. 
We also show the basic proof for the proposed functions.

\begin{figure*}[!h]
\centering
\includegraphics[width=0.6\textwidth]{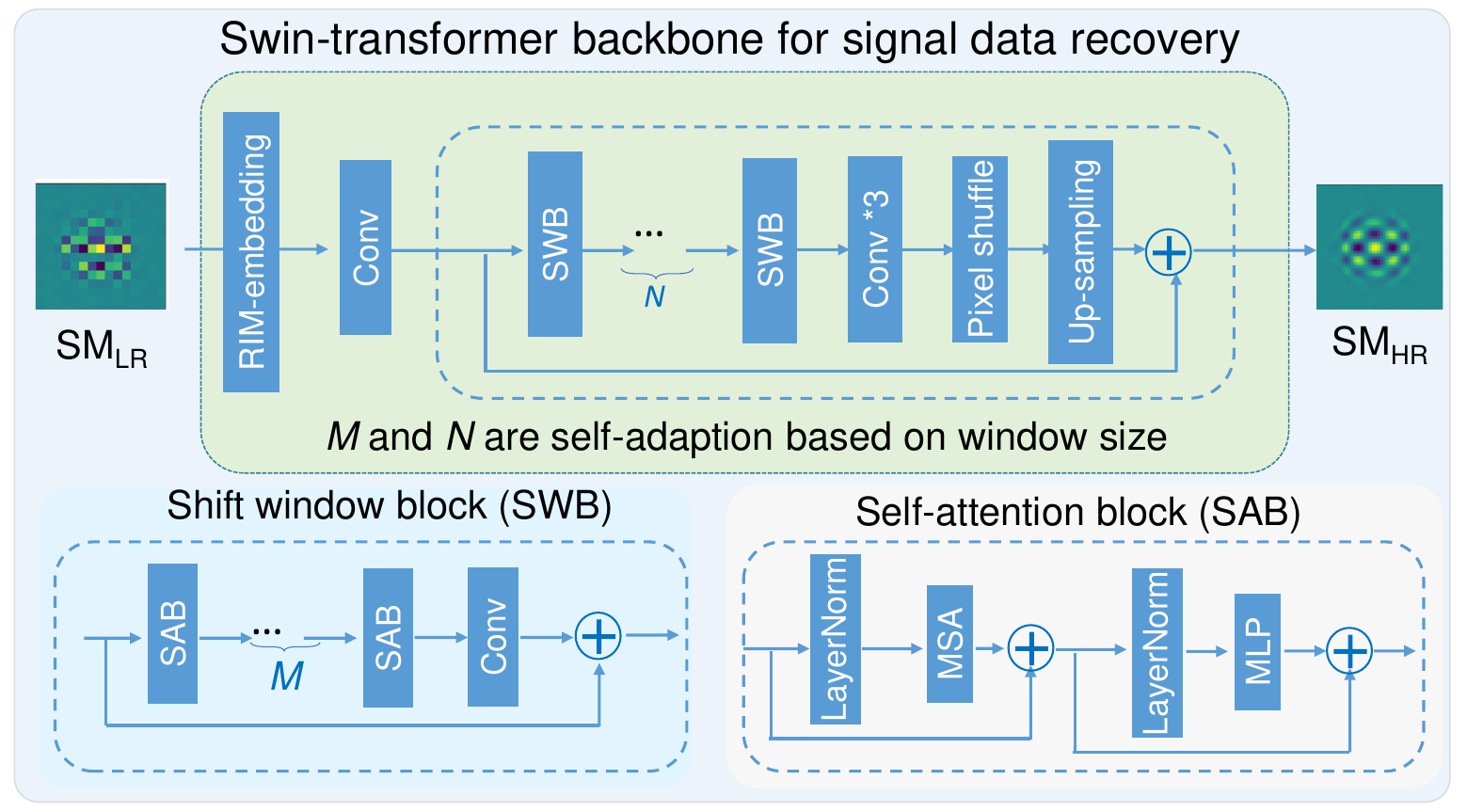}
\caption{Architecture of our proposed network for high-resolution signal matrix (${\text{SM}}_{{\text{HR}}}$) recovery.} 
\label{fig1-3net}
\end{figure*}

\textbf{Proof.} 
Based on the proposed formula, it is straightforward to prove that it satisfies the fundamental properties of a loss function, including non-negativity, differentiability, and convergence. The function ${L_{AD}}({p_i})$ is structured such that $\mu_{X_i}$ and $\mu_{Y_i}$ represent the mean features of datasets $X$ and $Y$, respectively. The Euclidean distance $\left\|\mu_{X_i}-\mu_{Y_i}\right\|_2$ measures the difference between these means, while the denominator provides a normalization factor, ensuring a reasonable scale for this difference. As the mean features $\mu_{X_i}$ approach $\mu_{Y_i}$, the numerator approaches zero, causing the sum to converge to zero and ${L_{AD}}({p_i})$ to approach 1, thereby minimizing the loss when predictions closely match actual values. The inclusion of the constant $c_1$ prevents division by zero, enhancing numerical stability. 

To prove that the defined loss function ${S_{AD}}({p_i})$ approaches zero under the condition that the covariance equals the product of the standard deviations, we first consider the definition of covariance. The covariance between two random variables $X$ and $Y$ is defined as $\operatorname{Cov}(X, Y)=$ $E\{[X-E(X)][Y-E(Y)]\}$. The correlation coefficient $\rho_{X Y}$ is given by the relationship between covariance and standard deviations, expressed as:
\begin{equation}
{\rho _{XY}} = \frac{{\operatorname{Cov} (X,Y)}}{{\sqrt {D(X)} \sqrt {D(Y)} }} = \frac{{{\sigma _{{X}{Y}}}}}{{{\sigma _{{X}}} \cdot {\sigma _{{Y}}}}},
\end{equation}  
where $D(X)$ and $D(Y)$ are the variances of $X$ and $Y$, respectively.

Next, we analyze the definition of the equation \ref{equalCov}. To ensure that the use of this loss function gradually leads the difference to converge to zero. Given $\sigma_{X Y}=$ $\sigma_{X} \cdot \sigma_{Y}$, this condition indicates that the covariance is equal to the product of the standard deviations, leading to a correlation coefficient $\rho_{X Y}=1$.

Finally, given as $\left|\rho_{X Y}\right|=1$, we can derive that the two images have a linear transformation relationship. Thus, we conclude that $E\left\{\left[Y-\left(a_0+b_0 X\right)\right]^2\right\}=0$. This implies that $0=$ $E\left\{\left[Y-\left(a_0+b_0 X\right)\right]^2\right\}=D\left[Y-\left(a_0+b_0 X\right)\right]+\left[E\left(Y-\left(a_0+b_0 X\right)\right)\right]^2$. Consequently, we arrive at $D\left[Y-\left(a_0+b_0 X\right)\right]=0$ and $E\left[Y-\left(a_0+b_0 X\right)\right]=0$. 

Indeed, in our specific cases, when our model perfectly matches the output, it indicates that the coefficients $a_0$ and $b_0$ take on the values $a_0=0$ and $b_0=1$. This means that the model can be expressed as $ Y=a_0+b_0 X=0+1 \cdot X=X $. This outcome signifies that the predicted values are identical to the actual values, demonstrating a perfect fit. In this case, the loss function  ${S_{AD}}({p_i})$ approaches zero, confirming that the model captures the underlying relationship without any error. Based on the definition and proof, we define our proposed SSIM absolute difference ($\textit{SSIM}_{AD}$) loss for SM recovery as:
\begin{equation}
\textit{SSIM}_{AD}({p_i}) = \frac{1}{{{N^2}}}\sum\limits_{{p_i} \in X,Y} {{L_{AD}}\left( {{p_i}} \right) \cdot {S_{AD}}\left( {{p_i}} \right)} ,
\end{equation}
where $N$ is the number of patches. It is important to note that due to the significant differences in the means and expectations across different regions of the entire signal map, using global means and variances as optimization targets cannot achieve the recovery of local details. Therefore, our patch-based loss function primarily focuses on enhancing the detail consistency in local area signal low- and high-frequency recovery. In our experiment, we found that training the model using only this function results in a slow convergence rate. To further address this limitation, inspired by work \cite{TCI_SSIM_Zhao,cvprross2017focal}, adding a factor to accelerate the training process combined this function with the $\ell_1$ norm. We formulate our proposed FSC-loss function as :
\begin{equation}
{\textit{SSIM}_{Mix}}(X,Y) = SSI{M_{AD}}(X,Y) \cdot {\ell_1},
\end{equation}
where $\ell_1$ is L1 Norm to learn global information and accelerate model training. We also experimentally demonstrate that choosing a simple but effective $\ell_1$ norm can balance the training speed (The training curves are shown in ablation study of Part IV, Fig. \ref{fig4losscompar}).

\subsection{SMSW Transformer}
 Inspired by \cite{liang2021swinir}, as shown in Figure \ref{fig1-3net}, we use a Swin-transformer backbone to construct a self-adaption multi-scale shift window (SMSW) Transformer for SM recovery. Our network contains three parts: 1) low-level feature extractor, 2) high-level feature extractor, and 3) ${\text{SM}}_{{\text{HR}}}$ recovery module. The low-level feature extractor consists of a convolution layer and the high-level extractor consists of four residual shift window Transformer blocks. The ${\text{SM}}_{{\text{HR}}}$ recovery module consists of a convolution layer and an up-sampling layer using pixel shuffle. Therefore, our network generates a ${\text{SM}}_{{\text{HR}}}$ by learning signal frequency distribution.

\section{Materials and Experiments}

\subsection{Dataset}
\subsubsection{Simulation Dataset without Noise}
We use same experimental setting \cite{shen2024simulation, TMIshigen10189221} for our simulation data. The simulation scanner models a magnetic particle imaging system based on field free point, and performs scans according to Lissajous trajectories. The simulation parameters are shown in Table \ref{tab:datapara}.

We set drive field signals with a frequency of 25 kHz and focus field signals with a frequency of 24.75 kHz are generated in the $(x, y)$ directions respectively to achieve two-dimensional scanning. The selection field gradient is kept consistent in both the $x$ and $y$ directions. Therefore, to form a fixed-size field of view (FOV), the drive field amplitudes $A_D$ and $A_E$ are calculated according to $ A_D=\frac{F O V_x \times G_x}{2}$ and $ A_E=\frac{F O V_y \times G_y}{2}$. We employ 16 signal signal matrixes for training, 4 for validation, and 5 for testing. 

\begin{table}[ht]
\renewcommand{\arraystretch}{1.1}
\setlength{\tabcolsep}{0.8pt}
\caption{Parameters for high resolution signal matrix in different datasets. Four system matrices in OpenMPI dataset are used with $\# \mathbf{7}, \# \mathbf{8}, \# \mathbf{9}$, and \#$\mathbf{10}$.} 
\begin{tabular}{cccc}
\hline Symbol & \begin{tabular}{l} 
Simulated \\
Dataset
\end{tabular} & \begin{tabular}{c} 
Open MPI \\
Dataset $(\# 7) /$ \\
$(\# \mathbf{8}, \# \mathbf{9}, \# \mathbf{1 0})$
\end{tabular} & \begin{tabular}{l} 
In-house \\
MPI Dataset
\end{tabular} \\
\hline FOV Size (mm) 
& $ 32 \times 32 \times 1$ & \begin{tabular}{c}
$37 \times 37 \times 18.5 /$ \\
$66 \times 66 \times 27$
\end{tabular} & \begin{tabular}{c} $40 \times 40 \times 1 /$ \\
$30 \times 30 \times 1$
\end{tabular} \\
\hline Grid Size $(p x)$ & $32 \times 32 \times 1$ & \begin{tabular}{c}
$37 \times 37 \times 37 /$ \\
$33 \times 33 \times 27$
\end{tabular} & $9 \times 9 \times 1$ \\
\hline Sequence & FFP-Lissajous & FFP-Lissajous & FFL / FFP \\
\hline MNP & {$[20.0-35.0] \mathrm{nm}$} & \begin{tabular}{c} 
Synomag / \\
Perimag
\end{tabular} & Perimag \\
\hline Sample Sz. (mm) & $1 \times 1 \times 1$ & $2 \times 2 \times 1$ & $2 \times 2 \times 1$ \\
\hline \begin{tabular}{l} 
\hline SF Grad. \\
$(\mathrm{T} / \mathrm{m})$
\end{tabular} & \begin{tabular}{c}
{$[2.0-4.0]$} \\
\end{tabular} & \begin{tabular}{c}
$-1 \times-1 \times 2 /$ \\
$-0.5 \times-0.5 \times 1$
\end{tabular} & \begin{tabular}{c}
{$0.6$}\\
$-1.7 \times-1.7 \times 3.4$
\end{tabular} \\
\hline DF Freq. & $25.0 \mathrm{kHz}$ & \begin{tabular}{c}
$2.5 / 102 \mathrm{MHz}$ \\
$2.5 / 96 \mathrm{MHz}$ \\
$2.5 / 99 \mathrm{MHz}$
\end{tabular} & $25.1 \mathrm{kHz}$ \\
\hline DF Amp. (mT) & $0 \times 0 \times[32-64]$ & $12 \times 12 \times 12$ & 12mT\\
\hline
\end{tabular}
\label{tab:datapara}
\end{table}

\subsubsection{Simulation Dataset with Noise}
To better simulate the real acquisition process of the SM, we added Gaussian noise with a signal-noise-ratio (SNR) of 20 dB to the ideal simulation process while keeping other default parameters consistent.

\subsubsection{Open MPI Dataset with Perimag Tracer}
Our proposed method is evaluated in Open MPI Data \cite{RN882openMPIdata}. The Open MPI Data is a public dataset based on SM reconstruction (Bruker, Ettlingen) \cite{knopp2016mdf}. This data contain MPI image signals and SMs from multiple repeated measurements of different magnetic particles. To ensure a fair comparison of experimental results, the main settings of this article are kept consistent with the previously published literature,  \cite{TMI_TranSMS}. This experiment includes three SMs for the construction of the training set (\#8, \#9) and the testing set \#10), which are measured based on Perimag (Micromod GmbH, Germany) particles. Signals were collected for the three-dimensional coordinates of the coil at a rate of 2.5 MS/s. Following ref  \cite{TMI_TranSMS}, frequency components with a signal-to-noise ratio (SNR) of greater than 5 were chosen. To facilitate more efficient model learning and save computational resources, we process the three-dimensional SM into separate two-dimensional SM slices.
\subsubsection{Open MPI Dataset with Synomag-D Tracer}
To validate the model's generalization performance, we conducted experiments using another SM based on Synomag-D Tracer. Unless otherwise specified, other parameters remain consistent with the publicly available dataset mentioned above. In our study, we treat SM recovery as a super-resolution task, dataset \#$\mathbf{7}$ with 101861 images, Dataset \#$\mathbf{8}$ with 65340 images, Dataset \#$\mathbf{9}$ with 68278 images, and Dataset \#$\mathbf{10}$ with 82485 images (we treat each frequency in SM as an image according to grid size, more details can be found in the work \cite{TMI_TranSMS}).
\subsubsection{In-house Dataset for application test}

Based on the trained model, we conducted application testing of its performance on three in-house developed MPI systems.
\begin{itemize}
    \item \textbf{In-house MPI system 1} \cite{TIM_yin2023streamlined}. This MPI system is based on field-free point (FFP) scanners. The selection field gradients are set as $[-1.7, -1.7, 3.4]$ T/m along the $X$, $Y$, and $Z$ axes.  A Cartesian trajectory was employed for scanning the field of view (FOV). More details are listed in Table \ref{tab:datapara}.

    \item \textbf{In-house MPI system 2} \cite{TBE-liguanghui}.  This MPI system is based on field-free line (FFL) scanners. The selection field gradients are set as 0.6 $\mathrm{T}/\mathrm{m}$ along the $X$ axis, with a drive frequency of 2.51 $\mathrm{kHz}$. During 2D imaging, the object rotates around the Z-axis within the FOV. A square grid of $9 \times 9$ was used for the SM measurement with a delta sample of $3 \times 3 \mathrm{mm}^2$. SM is with a grid size of $40 \times 40$.
    
    \item \textbf{In-house MPI system 3} \cite{TMI_hejie}. This in-house MPI system is an open-sided MPI system. The SM is measured within a cylindrical FOV with a grid size of $N_r \times 2 N_\theta \times N_z$. Sequential scan-based original cylindrical-FOV SMC requires $2 N_r N_\theta^2 N_z^2$ measurements. 

\end{itemize}

\subsection{Experimental Setup}
Our network is implemented in PyTorch and trained on our computing platform with a four 32G V100 GPUs. We empirically set learning rate of $10^{-1}-10^{-4}$ for different scales, and the AdamW optimizer is used for optimization with default parameters. We adopt a cosine learning rate schedule. The iterations are set as $3\cdot10^5$ and batch size is 128. Unless otherwise specified, for fair comparison, our setting is consistent with the settings for data partition in Alper's article \cite{TMI_TranSMS}.

\subsection{Competing Methods}
Following the recent related work for SM recovery \cite{TMI_TranSMS}, we evaluate our method against competing methods including: 

\begin{itemize}
    \item \textbf{Bicubic Interpolation} \cite{gungor2020super}. Bicubic and Strided Bicubic are widely-used interpolation methods for super-resolution tasks.
    \item \textbf{Compressed Sensing (CS)} \cite{weber2015reconstructionCS,8630849}.  The CS leverages the sparsity of signals, enabling the reconstruction of signals with far fewer measurements. It allows signal reconstruction from sub-Nyquist rate samples using optimization algorithms.
    \item \textbf{SRCNN} \cite{dong2015imageSRCNN}. A shallow convolutional network (three layers) for natural image super-resolution task. 
    \item \textbf{ VDSR} \cite{kim2016accurateVDSR}.A deep CNN (20 layers) for super-resolution task, which employs the residual network to model the relationship for low- and high-resolution images and avoid the gradient vanishing and explosion problem during training.
     \item \textbf{2d-SMRnet} \cite{3dsmr}. A CNN-based network for 2D SM recovery, which is revised from an existing method of 3d-SMRnet for SM recovery.
    \item \textbf{TranSMS} \cite{TMI_TranSMS}. A state-of-the-art transformer-based network for 2D SM recovery. The proposed network is the first method that combines the merits of CNN and Transformer for SM recovery and shows encouraging performance for the SM recovery as a super-resolution task. 
    \item \textbf{ProTSM} \cite{TMIshigen10189221}. This is our privious work for 3D SM recovery. We proposed a transformer-based method to exploit the physical characteristics of the SM coil channel and frequency index as prior information. For fair comparision, we compare this method on a same public dataset.
\end{itemize}

The performance of the methods is evaluated via calculating the normalized root mean squared error (nRMSE) for each frequency component.
\subsection{Quantitative Assessments}

We evaluate the performance of different methods in ${\text{SM}}_{{\text{HR}}}$ calibration and image reconstruction using nRMSE and peak Signalto-Noise-Ratio (pSNR). Assuming $\hat{\mathbf{A}}$ is the calibrated ${\text{SM}}_{{\text{HR}}}$ and A is the ground truth of $\hat{\mathbf{A}}$, the nRMSE can be defined as:

\begin{equation}
\textit{nRMSE}(\hat{\mathbf{A}})=\|\hat{\mathbf{A}}-\mathbf{A}\|_{F} /\|\mathbf{A}\|_{F},
\end{equation}
where $\|\mathbf{\bullet}\|_{F}^{2}$ is the Frobenius norm of the residual error and reference matrix.
For quality evaluation of reconstructed image, we employ the widely used pSNR. The pSNR is defined as:
\begin{equation}
\textit{pSNR}({\text{X}}) = 20{\log _{10}}\left( {\frac{{\sqrt K {{\left\| {{{\text{X}}_{ref}}} \right\|}_\infty }}}{{{{\left\| {\widetilde {\text{X}} - {{\text{X}}_{{\text{ref }}}}} \right\|}_2}}}} \right),
\end{equation}
where $K$ is the total number of pixels, and $\mathbf{X}_{ref}$ and $\widetilde {\text{X}}$ are the ground truth of reference image and predicted image, respectively.
\begin{table}[h]
    \renewcommand{\arraystretch}{1.2}
    \setlength{\tabcolsep}{8pt}
    \centering
    \caption{Average nRMSE (\%) in high-resolution signal matrix (${\text{SM}}_{{\text{HR}}}$) recovery for simulation Data (no noise) at scales of 2×-8×.}
    \resizebox{0.9\linewidth}{!}{ 
    \begin{tabular}[c]{c|c|c|c}
    \hline \multirow{2}{*}{Methods } & \multicolumn{3}{c}{ Scale factor } \\
     \cline { 2 - 4 } & $2 \times$ & $4 \times$ & $8 \times$\\
    \hline Bicubic & $12.29 \%$ & $25.28 \%$ & $54.59 \%$ \\
    \hline CS \cite{weber2015reconstructionCS} & $18.33 \%$ & $21.19 \%$& $90.26 \%$ \\
    \hline SRCNN \cite{dong2015imageSRCNN} & $14.82 \%$ & $54.44\%$ & $85.67 \%$ \\
    \hline VDSR \cite{kim2016accurateVDSR} & $17.71 \%$ & $42.78 \%$ & $123.02 \%$ \\
    \hline 2d-SMRnet \cite{3dsmr}  & $6.31 \%$ & $28.38 \%$ & $49.35 \%$ \\
    \hline TranSMS \cite{TMI_TranSMS} &$7.02\%$ & $31.8\%$  & $ 97.33\%$ \\
    \hline Ours & $\mathbf{5.97\%}$  & $\mathbf{10.19\%}$  & $\mathbf{28.57\%}$\\
    \hline
    \end{tabular}}
    \label{modelcompSimulation_No_noise} 
\end{table}

%
\begin{table}[ht!]
    \renewcommand{\arraystretch}{1.2}
    \setlength{\tabcolsep}{8pt}
    \caption{Average nRMSE (\%) in high-resolution signal matrix (${\text{SM}}_{{\text{HR}}}$) recovery for simulation Data with 20 db noise at scales of 2×-8×.}
    \label{modelcompSimulation_noise} 
    \centering
    \resizebox{0.9\linewidth}{!}{ 
    \begin{tabular}[c]{c|c|c|c}
    \hline \multirow{2}{*}{Methods } & \multicolumn{3}{c}{ Scale factor } \\
     \cline { 2 - 4 } & $2 \times$ & $4 \times$ & $8 \times$\\
    \hline Bicubic & $14.90 \%$ & $34.32 \%$ & $58.27 \%$ \\
    \hline CS  \cite{weber2015reconstructionCS} & $20.57 \%$ & $30.12 \%$ & $98.95\%$ \\
    \hline SRCNN \cite{dong2015imageSRCNN} & $22.29 \%$  & $49.65 \%$ & $75.11 \%$ \\
    \hline VDSR \cite{kim2016accurateVDSR} & $33.15 \%$ & $ 50.29\%$ & $75.25 \%$ \\
    \hline 2d-SMRnet \cite{3dsmr}  & $ 12.51\%$ & $35.12 \%$ & $ 101.79\%$ \\
    \hline TranSMS \cite{TMI_TranSMS} &$13.50\%$ & $39.50\%$ & $91.90\%$ \\
    \hline Ours & $\mathbf{7.75\%}$ & $\mathbf{15.04\%}$& $\mathbf{57.90\%}$ \\
    \hline
    \end{tabular}}
\end{table}

\begin{figure*}[!th]
\centering
\includegraphics[width=0.85\textwidth]{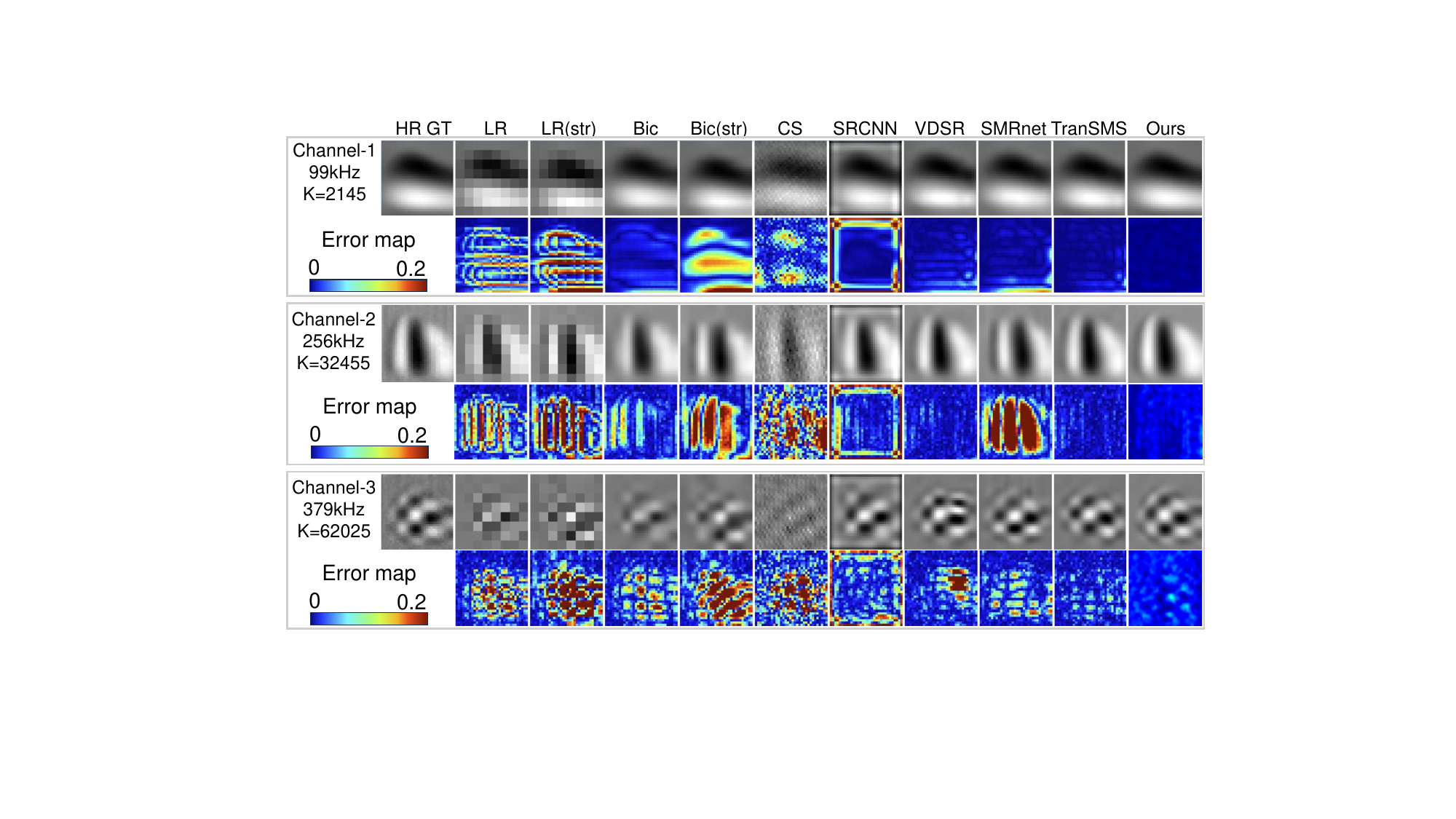}
\caption{Error maps of different methods for high resolution SM recovery in Open MPI with four times downsampling. K: respective indexes of different frequencies. HR GT: high resolution ground truth. LR: low resolution.} 
\label{fig2}
\end{figure*}

\section{Results and Discussion}
We evaluate different methods on five datasets. Our experimental results consist of two parts: 1) Evaluating all methods for SM recovery under different super-resolution factors on simulated datasets; 2) Further evaluating all methods on four real public MPI datasets to demonstrate its practical usability and generalization capabilities. 3) Application in our three in-house MPI systems.

\subsection{Result on Simulation Dataset}
As shown in Table \ref{modelcompSimulation_No_noise} and \ref{modelcompSimulation_noise}, all methods are evaluated in two simulation test sets (without or with noise). Our methods outperform the SOTA methods for 2×-8× scales. Furthermore, compared to competing method of TranSMS, our method yields superior performance against it, with more than 15\% lower nRMSE for downsampling scales of 4× and 8×. 

\subsection{Result on Open MPI Dataset}
As shown in Table \ref{modelcompReal8and9}, our methods outperform the SOTA methods for 4× and 8× scales. Furthermore, compared to competing method of TranSMS, our method yields superior performance against it, with 1-2\% lower nRMSE for downsampling scales of 4× and 8×. For 2× downsampling, considering the rich information already present in the input for this task, the error between the interpolation (Bicubic) result and ground truth (GT) is only 4.55\%. Therefore, with limited room for error reduction, our method achieves the same results as TranSMS (3.15\%) at down-sampling scale of 2× for input SM. However, as shown in Table \ref{modelcompreal7}, our model outperforms all competing methods when tracer is different in training and test sets.
\begin{table}[h!]
    \renewcommand{\arraystretch}{1.2}
    \setlength{\tabcolsep}{8pt}
    \caption{Average nRMSE (\%) in high-resolution signal matrix (${\text{SM}}_{{\text{HR}}}$) recovery for Open MPI Data at scales of 2×-8×. Str.: Strided Bicubic.}
    \centering
    \resizebox{0.8\linewidth}{!}{ 
    \begin{tabular}[c]{c|c|c|c}
    \hline \multirow{2}{*}{Methods } & \multicolumn{3}{c}{ Scale factor } \\
     \cline { 2 - 4 } & $2 \times$ & $4 \times$ & $8 \times$\\
    \hline Bicubic & $4.55 \%$ & $18.13 \%$ & $52.02 \%$ \\
    \hline Bicubic (str.) & $16.86 \%$ & $47.41 \%$ & $92.08 \%$ \\
    \hline CS \cite{weber2015reconstructionCS} & $8.81 \%$ & $51.48 \%$ & $101.31 \%$ \\
    \hline SRCNN \cite{dong2015imageSRCNN} & $53.65 \%$ & $53.16 \%$ & $79.29 \%$ \\
    \hline VDSR \cite{kim2016accurateVDSR} & $4.55 \%$ & $12.62 \%$ & $39.19 \%$ \\
    \hline 2d-SMRnet \cite{3dsmr}  & $19.69 \%$ & $48.94 \%$ & $75.37 \%$ \\
    \hline TranSMS \cite{TMI_TranSMS} &$3.15\%$ & $6.19\%$ & $20.58 \%$ \\
    \hline Ours & $\mathbf{3.15\%}$ & $\mathbf{4.13 \%}$& $\mathbf{14.87 \%}$\\
    \hline
    \end{tabular}
    }
    \label{modelcompReal8and9} 
\end{table}

\begin{table}[h!]
    \renewcommand{\arraystretch}{1.2}
    \setlength{\tabcolsep}{8pt}
    \caption{Average nRMSE (\%) in (${\text{SM}}_{{\text{HR}}}$) recovery for OpenMPI Data (generalized investigation for different tracers, training with Synomag-D and testing with Perimag).}
    \centering
    \resizebox{0.8\linewidth}{!}{ 
    \begin{tabular}[c]{c|c|c|c}
    \hline \multirow{2}{*}{Methods } & \multicolumn{3}{c}{ Scale factor } \\
     \cline { 2 - 4 } & $2 \times$ & $4 \times$ & $8 \times$\\
    \hline Bicubic & $4.55 \%$ & $18.13 \%$ & $52.02 \%$ \\
    \hline Bicubic (str.) & $16.86 \%$ & $47.41 \%$ & $92.08 \%$ \\
    \hline CS \cite{weber2015reconstructionCS} & $8.81 \%$ & $51.48 \%$ & $101.31 \%$ \\
    \hline SRCNN \cite{dong2015imageSRCNN} & $50.88 \%$ & $62.81 \%$ & $106.76 \%$ \\
    \hline VDSR \cite{kim2016accurateVDSR} & $3.34 \%$ & $11.83 \%$ & $113.81 \%$ \\
    \hline 2d-SMRnet \cite{3dsmr}  & $6.85 \%$ & $17.22 \%$ & $78.88 \%$ \\
    \hline TranSMS \cite{TMI_TranSMS} &$3.32\%$ & $10.66\%$ & $ 114.45\%$ \\
    \hline ProTSM \cite{TMIshigen10189221}& $\mathbf{3.13\%}$ & $9.88\%$ & $49.98\%$\\
    \hline Ours & $3.29\%$ & $\mathbf{8.10 \%}$ & $\mathbf{ 37.09\%}$\\
    \hline
    \end{tabular}}
    \label{modelcompreal7} 
\end{table}

 As shown in Figure \ref{fig2}, error maps and recovered ${\text{SM}}_{{\text{HR}}}$ recovery for three frequencies by different methods at 4× SR are displayed. Our method reconstructs sharper ${\text{SM}}_{{\text{HR}}}$ than competing methods. Although the bicubic interpolation is easy-to-use, poor results are shown for high-frequency information recovery (379kHZ). The CS shows artifacts. The SRCNN exhibits blurry boundary artifacts. Although the TranSMS exhibits good performance 
 for sharp SM calibratrion at frequency (99kHZ and 256kHZ), there are details missing in high-frequency information responses (K=379kHZ). In contrast, our method yields an better results and exhibit less details missing in high-frequency information responses of sharp structure in high-frequency signal and similarity to the ground truth (lowest error).
\subsection{Image Reconstruction on Open MPI Dataset}
\begin{figure*}[h]
\centering
\includegraphics[width=\textwidth]{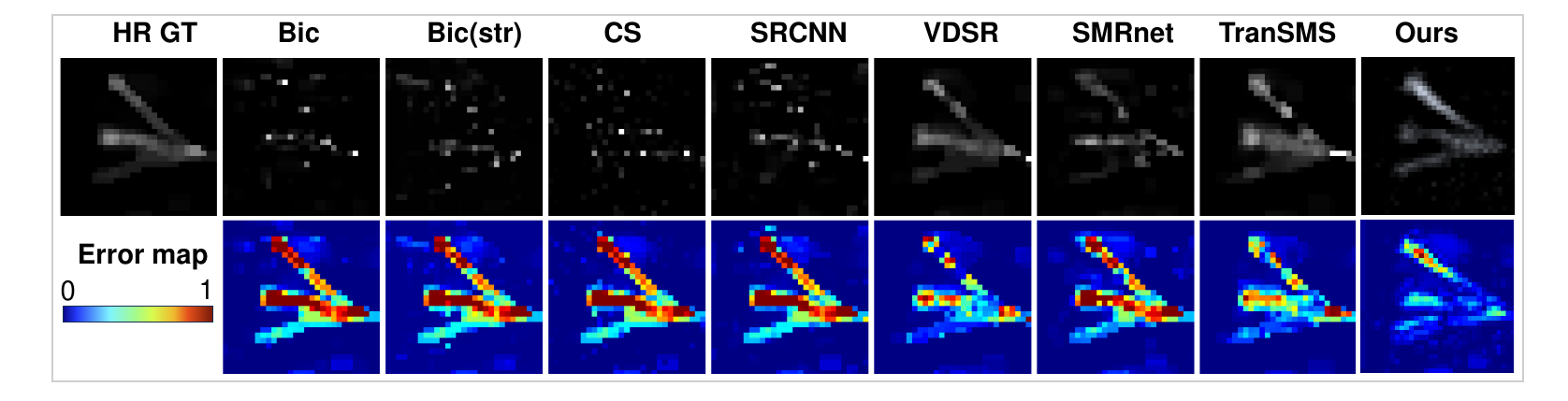}
\caption{Error maps of different methods for image reconstruction in Open MPI with four times downsampling. HR GT: high resolution ground truth.} 
\label{fig3image}
\end{figure*}

\begin{table}[h!]
    \renewcommand{\arraystretch}{1.2}
    \setlength{\tabcolsep}{8pt}
    \caption{Average pSNR (\%) in image reconstruction using predicted (${\text{SM}}_{{\text{HR}}}$) for Open MPI Data at scales of 2×-8×. the results for all methods are evaluated with same test set (\#10).}
    \centering
    \resizebox{0.8\linewidth}{!}{ 
    \begin{tabular}[c]{c|c|c|c}
    \hline \multirow{2}{*}{Methods } & \multicolumn{3}{c}{ Scale factor } \\
     \cline { 2 - 4 } & $2 \times$ & $4 \times$ & $8 \times$\\
    \hline Bicubic & $34.77 \%$ & $17.15 \%$ & $11.34 \%$ \\
    \hline Bicubic(str.) & $33.30 \%$ & $18.62 \%$ & $13.88 \%$ \\
    \hline CS \cite{weber2015reconstructionCS} & $33.89 \%$ & $18.62 \%$ & $13.88\%$ \\
    \hline SRCNN \cite{dong2015imageSRCNN} & $32.21 \%$ & $19.39\%$ & $- \%$ \\
    \hline VDSR \cite{kim2016accurateVDSR} & $38.23 \%$ & $30.59 \%$ & $10.01\%$ \\
    \hline 2d-SMRnet \cite{3dsmr}  & $33.79 \%$ & $24.81 \%$ & $12.60 \%$ \\
    \hline TranSMS \cite{TMI_TranSMS} &$38.54\%$ & $31.96\%$ & $13.38 \%$ \\
    \hline Ours & $\mathbf{38.91\%}$ & $\mathbf{ 33.62\%}$& $\mathbf{17.25\%}$\\
    \hline
    \end{tabular}}
    \label{tablemodelcompimage} 
\end{table}
As shown in the table \ref{tablemodelcompimage}, the results for image reconstruction demonstrate that our reconstructed images yield better results compared to the reference images, with the error visualization depicted in the Figure \ref{fig3image}.

\subsection{Ablation study}
\begin{figure}[htp]
\centering
\includegraphics[width=0.9\columnwidth]{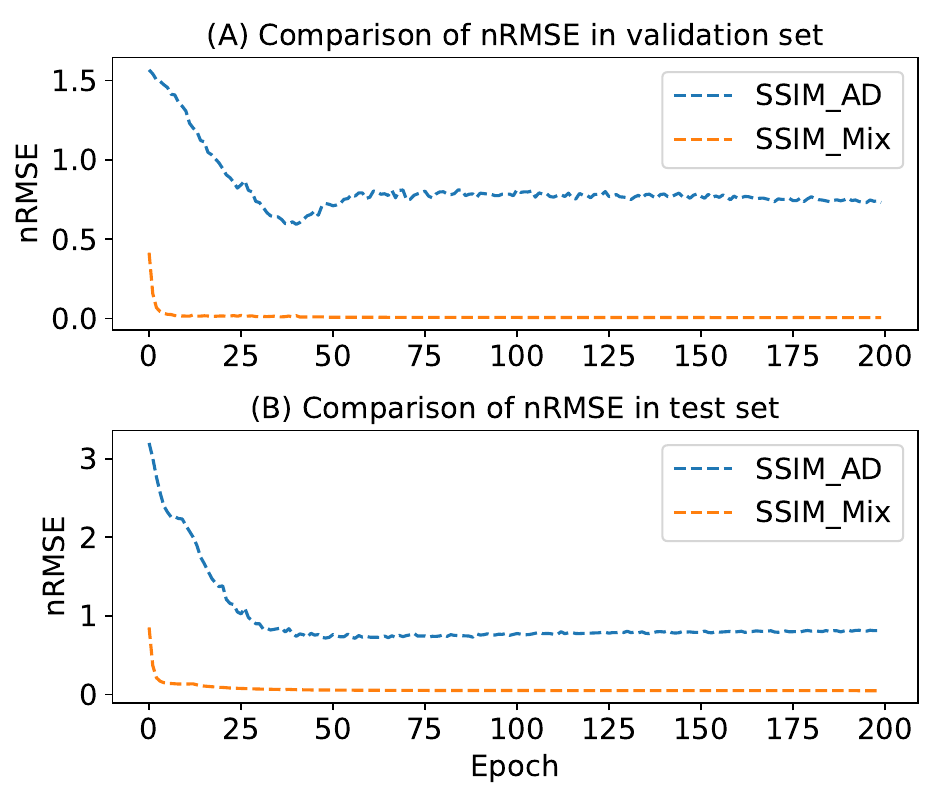}
\caption{Validation and test nRMSE against epoch for proposed loss.} 
\label{fig4losscompar}
\end{figure}

 As shown in Table \ref{tableablationstudy}, we also conduct ablation study for using existing loss to train our model. Our results confirm that the proposed loss function and RIM-embedding method can boost model performance. The results (Figure \ref{fig4losscompar}) also demonstrate that proposed loss function converges fast and shows better performance than $\textit{SSIM}_{AD}$ which is due to the only window-based calculation for local patches.

\begin{table}[ht]
    \renewcommand{\arraystretch}{1.2}
    \setlength{\tabcolsep}{8pt}
    \caption{Evaluation for each component of proposed approach and existing losses. SWSM: our self-adaption multi-scale shifted window-based (SMSW) network. FSC: proposed loss. RIM: proposed embedding method. SSIM: Structural similarity.}
    \centering
    \resizebox{0.8\linewidth}{!}{ 
    \begin{tabular}{c|c|c}
    \hline \multirow{2}{*}{Methods } & \multicolumn{2}{c}{ Scale factor } \\
    \cline { 2 - 3 } & $2 \times$ & $4 \times$ \\
    \hline SWSM+SSIM & $11.12 \%$ & $16.92 \%$ \\
    \hline SWSM+L1 & $3.37 \%$ & $6.83 \%$ \\
    \hline SWSM+L2 & $3.41 \%$ & $6.92 \%$ \\
    \hline SWSM+FSC & $3.20 \%$ & $6.0 \%$ \\
    \hline SWSM+RIM+SSIM & $10.05\%$ & $15.89\%$ \\
    \hline SWSM+RIM+L1 &$3.23\%$  &$5.76\%$  \\
    \hline SWSM+RIM+L2 & $3.32\%$ & $5.97\%$ \\
    \hline SWSM+RIM+FSC (Ours) & $\mathbf{3.15\%}$ & $\mathbf{4.14\%}$ \\
    \hline
    \end{tabular}}
    \label{tableablationstudy} 
\end{table}
\subsection{Application in Our In-house MPI Systems}

\begin{figure}[h]
\centering
\includegraphics[width=0.9\columnwidth]{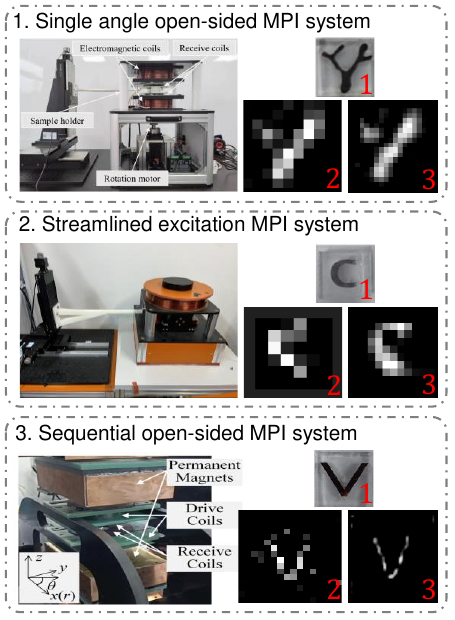}
\caption{Application in three in-house MPI systems. The number 1, 2, and 3 denote phantom, reconstructed image with measured raw poor quality SM, and reconstructed image with predicted high-resolution SM, respectively.} 
\label{fig5application}
\end{figure}

As shown in Figure \ref{fig5application}, By integrating our algorithm into our three in-house MPI systems \cite{TIM_li2024modified,TIM_yin2023streamlined,TMI_hejie}, we achieved performance enhancements: 1) Reduced acquisition time: Our algorithm cuts acquisition time by over fourfold, eliminating the need for costly high-resolution SMs. 2) Improved imaging resolution: The algorithm enhances image clarity and resolution, overcoming the limitations of low-resolution SM and enabling clear differentiation of phantom shapes. This leads to reduced labor costs in practical applications, highlighting the algorithm's effectiveness for improving MPI system performance.

\section{Conclusion}
We proposed a novel method to achieve accurate SM recovery and reduce the calibration time of ${\text{SM}_{\text{HR}}}$. The results demonstrate that our learning approach outperforms existing SOTA methods for SM recovery. Our method also is a useful tool for application in our in-house MPI systems to improve reconstructed image quality and shorten SM calibration time.
%
%
%

\printbibliography

%

\end{document}